\documentclass[preprint,showpacs,superscriptaddress,preprintnumbers,amsmath,amssymb]{revtex4}

\usepackage{graphicx}
\usepackage{dcolumn} % Align table columns on decimal point
\usepackage{bm} % bold math
\usepackage[figuresright]{rotating}

\newcommand{\BE}{$B$($E2$)}
\newcommand{\E}{$E$(2$_1^+$)}
\newcommand{\Ca}{$^{16}$C}
\newcommand{\Cb}{$^{18}$C}

\begin{document}

\title{Lifetime measurements of first excited states in $^{16,18}$C }

\author{H.~J.~Ong}
\email{onghjin@ribf.riken.jp}
\thanks{Present address: RIKEN Nishina Center, Hirosawa 2-1,
  Wako, Saitama 351-0198, Japan}
\affiliation{Department of Physics, University of Tokyo, Hongo 7-3-1,
              Bunkyo, Tokyo 113-0033, Japan }
\author{N.~Imai}
\thanks{Present address: KEK, Oho 1-1, Tsukuba, 
  Ibaraki 305-0801, Japan}
\affiliation{RIKEN Nishina Center, Hirosawa 2-1, Wako, Saitama 351-0198, Japan}
\author{D.~Suzuki}
\affiliation{Department of Physics, University of Tokyo, Hongo 7-3-1,
              Bunkyo, Tokyo 113-0033, Japan }
\author{H.~Iwasaki}
\thanks{Present address: Institut f$\ddot{\rm u}$r Kernphysik,
  Universit$\ddot{\rm a}$t zu K$\ddot{\rm o}$ln, Germany}
\affiliation{Department of Physics, University of Tokyo, Hongo 7-3-1,
              Bunkyo, Tokyo 113-0033, Japan }
\author{H.~Sakurai}
\thanks{Present address: RIKEN Nishina Center, Hirosawa 2-1,
  Wako, Saitama 351-0198, Japan}
\affiliation{Department of Physics, University of Tokyo, Hongo 7-3-1,
              Bunkyo, Tokyo 113-0033, Japan }
\author{T.~K.~Onishi}
\affiliation{Department of Physics, University of Tokyo, Hongo 7-3-1,
              Bunkyo, Tokyo 113-0033, Japan }
\author{M.~K.~Suzuki}
\affiliation{Department of Physics, University of Tokyo, Hongo 7-3-1,
              Bunkyo, Tokyo 113-0033, Japan }
\author{S.~Ota}
\affiliation{
  Center for Nuclear Study,
              University of Tokyo,
              RIKEN campus, Hirosawa 2-1, Wako, Saitama 351-0198, Japan}
\author{S.~Takeuchi}
\affiliation{RIKEN Nishina Center, Hirosawa 2-1, Wako, Saitama 351-0198, Japan}
\author{T.~Nakao}
\affiliation{Department of Physics, University of Tokyo, Hongo 7-3-1,
              Bunkyo, Tokyo 113-0033, Japan }
\author{Y.~Togano}
\affiliation{Department of Physics, Rikkyo University,
        Nishi-Ikebukuro 3-34-1, Toshima, Tokyo 171-8501, Japan}
\author{Y.~Kondo}
\thanks{Present address: RIKEN Nishina Center, Hirosawa 2-1,
  Wako, Saitama 351-0198, Japan}
\affiliation{Department of Physics, Tokyo Institute of Technology, 
	Ookayama 2-12-1, Meguro, Tokyo 152-8551, Japan}
\author{N.~Aoi}
\affiliation{RIKEN Nishina Center, Hirosawa 2-1, Wako, Saitama 351-0198, Japan}
\author{H.~Baba}
\affiliation{RIKEN Nishina Center, Hirosawa 2-1, Wako, Saitama 351-0198, Japan}
\author{S.~Bishop}
\affiliation{RIKEN Nishina Center, Hirosawa 2-1, Wako, Saitama 351-0198, Japan}
\author{Y.~Ichikawa}
\affiliation{Department of Physics, University of Tokyo, Hongo 7-3-1,
              Bunkyo, Tokyo 113-0033, Japan }
\author{M.~Ishihara}
\affiliation{RIKEN Nishina Center, Hirosawa 2-1, Wako, Saitama 351-0198, Japan}
\author{T.~Kubo}
\affiliation{RIKEN Nishina Center, Hirosawa 2-1, Wako, Saitama 351-0198, Japan}
\author{K.~Kurita}
\affiliation{Department of Physics, Rikkyo University,
        Nishi-Ikebukuro 3-34-1, Toshima, Tokyo 171-8501, Japan}
\author{T.~Motobayashi}
\affiliation{RIKEN Nishina Center, Hirosawa 2-1, Wako, Saitama 351-0198, Japan}
\author{T.~Nakamura}
\affiliation{Department of Physics, Tokyo Institute of Technology,
        Ookayama 2-12-1, Meguro, Tokyo 152-8551, Japan}
\author{T.~Okumura}
\affiliation{Department of Physics, Tokyo Institute of Technology,
  Ookayama 2-12-1, Meguro, Tokyo 152-8551, Japan}
\author{Y.~Yanagisawa}
\affiliation{RIKEN Nishina Center, Hirosawa 2-1, Wako, Saitama 351-0198, Japan}

\date{\today}% It is always \today, today,
     %  but any date may be explicitly specified

\begin{abstract}
%%%%%%%%0123456789012345678901234567890123456789012345678901234567890123456789
	The electric quadrupole transition from the first 2$^+$ state
	to the ground 0$^+$ state in $^{18}$C was studied through
	lifetime measurement by an upgraded recoil shadow method applied
	to inelastically scattered radioactive $^{18}$C nuclei.  The
	measured mean lifetime is $18.9 \pm 0.9 \ ({\rm
	stat}) \pm 4.4 \ ({\rm syst})$~ps, corresponding to a
	$B$($E2;2_1^+\rightarrow 0^+_{\rm gs}$) value of $4.3\pm 0.2 \pm 1.0 \
	e^2$fm$^4$, or about 1.5 Weisskopf units.  The mean lifetime
	of the first 2$^+$ state in $^{16}$C was remeasured to be
	$18.0 \pm 1.6 \pm 4.7 $~ps, about four times shorter than the value
	reported previously.  The discrepancy between the two results
	was resolved by incorporating the $\gamma$-ray angular distribution
	measured in this work into the previous measurement.
        These transition strengths are hindered compared to the
	empirical transition strengths, indicating that the anomalous
	hindrance observed in $^{16}$C persists in $^{18}$C.
\end{abstract}

\pacs{23.20.Js, 21.10.Tg, 29.30.Kv} % PACS, the Physics and Astronomy
     % Classification Scheme.
\maketitle

\section{Introduction}

        Structure of neutron-rich nuclei is one of the frontiers in
	nuclear physics. In such nuclei, several exotic
	phenomena like a halo~\cite{halo} or skin~\cite{skin}
	structure and disappearance of the traditional magic
	numbers~\cite{iwasaki,moto,scheit} have been discovered with
	the advance of experimental techniques and accelerators.
	These findings are unexpected by the standard nuclear
	structure model and precede the development of the theory.

	Recently, we reported another exotic phenomenon of extremely
	suppressed \BE\ value for the transition between the
	first $2^+$ ($2^+_1$) state to the ground ($0^+_{\rm gs}$) state
	in neutron-rich \Ca~\cite{imai}.  The \BE\ was obtained by measuring
	the mean lifetime of the $2^+_1$ state  ($\tau(2^+_1)$) using a new
	experimental technique.  In general, an even-even atomic
	nucleus has properties as a quantum liquid drop.
	In the liquid-drop model, the \BE\ value is inversely
	proportional to the excitation energy of the $2^+_1$ state
	(\E)~\cite{be2}.
	However, the measured \BE\ of \Ca\ was found to deviate greatly
	from the value expected by the empirical formula~\cite{raman} of
	the \E,  indicating a suppressed proton collectivity in \Ca.

	Contrary to the suppressed proton collectivity, a large
	neutron collectivity was suggested based on the measurement
	of the interference between the nuclear and electromagnetic
	interactions in the excitation from the ground state to the
	$2^+_1$ state~\cite{elekes}.  Meanwhile, the large quadrupole
	deformation length of \Ca\ extracted from the proton inelastic
	scattering also indicates that the neutrons contribute
	predominantly to the strength of the excitation to the
	$2^+_1$ state, whereas the protons seem to be frozen~\cite{onghjin}.

        The suppressed \BE\ may indicate quenched effective charges and/or
        the emergence of a new magic number $Z=6$ in the
        light neutron-rich carbon isotopes.  For neutron-rich nuclei,
        the isovector quadrupole mode is likely to be enhanced owing to
        the large isospin.  Such enhancement will give rise to quenched
	isovector effective charges~\cite{BM,sagawa-asahi}, in particular,
	a small neutron effective charge that reduces the contribution of
	the valence neutron(s) to the \BE\ value.
	Indeed, the quenched effective charges have been observed 
	in the neighboring $^{15,17}$B nuclei~\cite{15B,17B}.
	Moreover, in the case of the closed shell nuclei, the isoscalar
	polarization will also be reduced~\cite{BM}.
	In this light, the emergence of the proton magic number
	$Z=6$ has been suggested by a shell model calculation~\cite{fuji}.

	Besides the shell model calculation, several other microscopic
	models have also been proposed to explain the
	mechanism of the small \BE\ value for \Ca.  A calculation
	using antisymmetrized molecular dynamics (AMD) has attributed
	the anomalous feature to the opposite deformations in the
	proton and the neutron matters~\cite{amd1,amd2}.  It is
	interesting to note
	that calculations assuming a simple structure that $^{16}$C is
	composed of $^{14}$C +$n$+$n$ have also reproduced the \E\ and
	the \BE\ of this nucleus~\cite{suzuki1,suzuki2,hagino}.
	To shed light on the exotic phenomenon exhibited by \Ca\
	and to scrutinize the claim for the emergence of the $Z=6$
	magic number in the neutron-rich C isotope, more experimental
	information specifically the information on the neighboring
	\Cb\ isotope is needed.

	In this article, we report the lifetime measurement of the
	$2^+_1$ state in the \Cb\ nucleus by means of an
	upgraded recoil shadow method.  The
	remeasurement of the lifetime of the $2^+_1$ state in \Ca\ is
	also reported.
	The $\tau(2^+_1)$ of \Cb\ is expected to be as long
	as that for \Ca\ since the energy level schemes of the nuclei
	are almost identical.  Specifically, the \E\ values are
	1766~\cite{ToI} and 1585(10)~keV~\cite{18clevel}; the one-neutron
	separation energies ($S_n$) are 4250(4) and 4180(30)~keV,
	for \Ca\ and \Cb, respectively~\cite{ToI}.
	In addition, we have measured the lifetimes of
	the $1/2^-$ state in $^{11}$Be and the $3^-$ state in $^{16}$N
	to verify the method.

\section{Recoil Shadow Method (RSM)}
\label{sec_two}
        For earlier works on the recoil shadow method (RSM), we refer
	the readers to Ref.~\cite{limkilde,metag,bocquet,gueorguieva}.
	In both the previous~\cite{imai} and the present work, we applied
	the RSM to $\gamma$ decays from the excited radioactive nuclei.
	The RSM makes use of the shadow effect of a lead shield placed
	around a reaction target on NaI(Tl) detectors of a highly segmented
	$\gamma$-ray detector array.
	In this method, excited nuclei are produced via inelastic scattering
	or fragmentation reaction of an intermediate-energy radioactive
	nuclear beam (RNB) in the reaction target.  Each nucleus then travels
	a certain distance before decaying through emission of a
	$\gamma$ ray, which is detected by the NaI(Tl) detector
	array.  The yield of the $\gamma$ rays detected by
	each NaI(Tl) detector depends on the emission point and the
	$\gamma$-ray angular distribution; the latter is governed by the
	alignment produced by the nuclear reaction~\cite{diamond}.
	Since the flight length, i.e. the emission point of the de-excitation
	$\gamma$ ray depends on the velocity of the ejectile (which is known)
	and the lifetime of
	the excited state, assuming that we know the angular distribution of
	the $\gamma$ rays, the lifetime can be determined, in principle, by
	observing the yield distribution of the $\gamma$ rays.  For lifetimes
	of as short as a few tens of picoseconds, however, the yield
	distribution is not sensitive to the small variation of the emission
	point.  Hence, determination of such lifetimes
	is achieved with the presence of the lead shield, which enhances the
	lifetime dependence of the yield distribution.

	In the previous work~\cite{imai}, we bombarded a $^9$Be target with a
	40-MeV/nucleon \Ca\ beam and detected the de-excitation $\gamma$ rays
	with only two layers of NaI(Tl) detectors placed cylindrically
	around the beam axis. The mean lifetime was determined by
	comparing the ratio between the $\gamma$-ray yields detected in
	the two layers with a calculated ratio function, which was obtained
	for several mean lifetimes through simulations. However, the angular
	distribution of the $\gamma$ rays was not measured, and thus we
	had assumed the angular distribution obtained by a theoretical
	calculation.

	In this work, we have made three improvements to the RSM.
	First, we determined the lifetimes independent of the angular
	distribution.  For the determination of each lifetime,
	two measurements of the $\gamma$-ray yields were performed:
	one with the lead shield installed, and the other without
	the lead shield.  The $\gamma$-ray yield from the former
	measurement carries the information of both the lifetime and
	the $\gamma$-ray angular distribution, while that from the
	latter measurement carries mainly the information of the angular
	distribution.  Hence, we can determine the lifetime by
	comparing the ratio of the two $\gamma$-ray yields
	to a simulated ratio function.
        Second, we compensated the drop (by almost two orders of magnitude)
	in the beam intensity from \Ca\ to \Cb\ by increasing
	the number of detectors from 32 in the previous setup to 130.
	The last improvement was achieved by using a fast \Cb\ beam at
	an energy of about 79 MeV/nucleon, thus increasing the flight
	length by a factor of about 1.4.

	As will be shown later, the results for $^{11}$Be and $^{16}$N
	demonstrate that the present RSM is a reliable and powerful
	means to determine the lifetimes of nuclear excited states
	with simple or known $\gamma$-decay schemes.  Because the present
	RSM is independent of the $\gamma$-ray angular distribution, it 
	is applicable to any type of nuclear reaction that produces
	excited nuclei.  And as such, one can measure the lifetimes of
	excited states produced by any kind of reaction channels with the
	intermediate-energy RNB in a single experiment.  In fact, we
	have also successfully measured the lifetimes of excited
	states in $^{17}$C, produced by one-neutron knockout reaction
	of \Cb~\cite{dsuzuki}.

\section{Experiment}
\label{sec_exp}

        In the experiment, we measured the lifetime of the $2^+_1$
        state of \Cb\ populated via inelastic scattering of the \Cb\
        beam at 79~MeV/nucleon and those of the excited states in
        $^{11}$Be and $^{16}$N produced through breakup reaction of
        the \Cb\ beam.  We also measured the
        lifetime of the $2^+_1$ state of \Ca\ populated via two reactions,
        namely the inelastic scattering of a 72-MeV/nucleon \Ca\
        beam and the breakup reaction of the 79-MeV/nucleon \Cb\ beam.
        Furthermore, we measured the angular distribution of the $\gamma$
        rays for the $2^+_1$ state of \Ca\ inelastically excited at
        40~MeV/nucleon in order to incorporate it into the experiment data
	in Ref.~\cite{imai}.

	The experiment was performed at the RIKEN accelerator research
	facility.  Secondary beams of $^{16,18}$C were produced
	in two separate measurements 
	through projectile fragmentation of an 110-MeV/nucleon
	$^{22}$Ne primary beam, and separated by the RIPS beam
	line~\cite{rips}. Particle identification of the secondary
	beam was performed event-by-event by means of the
	time-of-flight (TOF)-$\Delta E$ method using two 1.0-mm-thick
	plastic scintillation counters located at the second and
	final focal planes of RIPS. The beams were directed at a 370-mg/cm$^2$
	$^{9}$Be target placed at the exit of the RIPS beam line.
	Two sets of parallel plate avalanche counters (PPACs) were placed
	upstream of the target to measure
	the position and angle of the projectile incident upon the
	target.  The $^{16,18}$C beam had typical intensities of
	$6.5\times 10^4$ and $ 2.3 \times 10^4$ particles per second,
	respectively.

	Outgoing particles from the target were identified by the
	$\Delta E$-$E$-TOF method using a plastic scintillator
	hodoscope~\cite{hodoscope} located 3.8~m downstream of the target.
	The hodoscope, with an active area of 1.0$\times$1.0~m$^2$,
	consisted of thirteen vertical $\Delta E$-scintillator slats 
	and sixteen horizontal $E$-scintillator bars with 5.0 mm and 
	60.0 mm thicknesses, respectively.  The scattering angles were
	determined by combining the hit position information on
	the hodoscope with the incident angles and hit positions on the
	target obtained by the PPACs.
 
\begin{figure}
\begin{center}
\includegraphics[width=80mm]{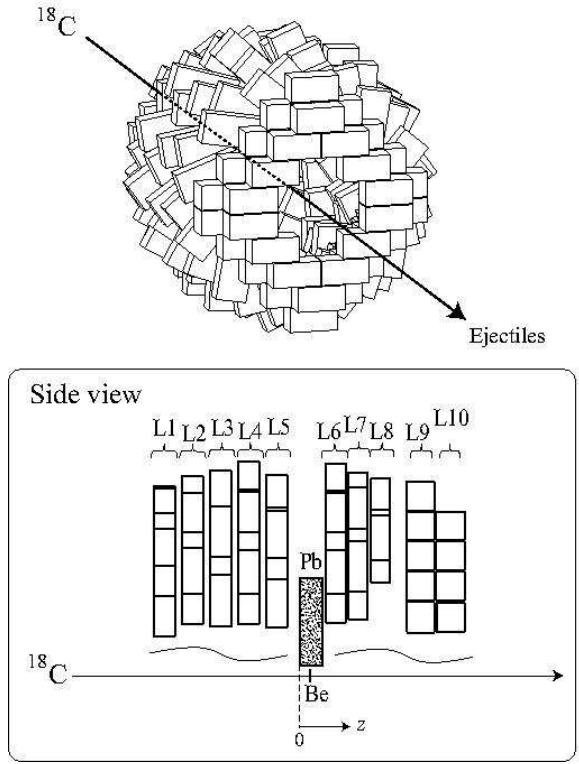}
\end{center}
\caption{Schematic view of $\gamma$-ray detectors. A beryllium target is
surrounded by a 5~cm-thick lead shield, and 10 layers of NaI(Tl)
scintillators (L1 $\sim$ L10) placed cylindrically around the beam axis.
For clarity, only part of the detectors and lead shield is shown in the
inset.}
\label{setup}
\end{figure}

	In order to implement the RSM concept, a thick $\gamma$-ray shield
	was placed around the target, as shown in Fig.~\ref{setup}.
	The shield was a 5~cm-thick lead block with an outer
	frame of 24$\times$24~cm$^2$ and an inner hole of
	5.4~cm in diameter. The inner hole surrounded the beam tube
	housing the $^{9}$Be target.  We removed the lead shield during 
	measurement of the angular distribution of $\gamma$ rays.  For
	the sake of later discussion, the $z$-axis is defined as the beam
	direction, which is close to the flight direction of the de-exciting
	nucleus. The origin of the $z$-axis, $z=0.00$~cm, was taken at the
	upstream edge of the lead shield.  We have carried out
	both measurements, namely the measurements with and without the
	lead shield, with the target placed at $z=-0.15$~cm and $z=2.05$~cm.
	We refer to these setups as the ``upstream setup'' and the
	``center setup'' hereinafter.

	The $\gamma$-rays from the excited nuclei in-flight were
	detected by an array of 130 NaI(Tl) detectors, which form part
	of the DALI1~\cite{dali1} and the DALI2~\cite{dali2}. The NaI(Tl)
	detectors, all of which are rectangular in shape, are of three
	different sizes:
	$4.5\times8\times16$~cm$^3$, $4\times8\times16$~cm$^3$,
	$6.1\times6.1\times12.2$~cm$^3$.  The array was divided into
	10~layers, labeled L1-10 as shown in Fig.~\ref{setup}, with
	each layer consisting of $10\sim 18$ detectors
	arranged coaxially with respect to the beam direction.
        The layers are closely packed to cover polar angular ranges
	of $15^{\circ}\sim 85^{\circ}$ and $100^{\circ}\sim 170^{\circ}$
	in the laboratory frame.  The detectors were mounted on and supported
	by nine 3-mm-thick aluminum plates.  The distances of the
	center of the detectors from the center of the Be target ranged
	from 28~cm to 56~cm.

	In the present work, we counted the number of full-energy-peak
	events detected by each layer during the measurements with and
	without the lead shield.  For convenience, the numbers obtained
	with the $i$-th layer during the former and the latter measurements
	are denoted by $N^i_{\rm wPb}$ and $N^i_{\rm woPb}$.  The
	$N^i_{\rm wPb}$ and $N^i_{\rm woPb}$ were obtained by fitting
	the measured $\gamma$-ray energy spectra with response
	functions obtained with simulations, plus $\gamma$-ray
	spectra for the $^{14}$C isotope as backgrounds.
	The spectra for $^{14}$C were selected because (i) all excited
	states in $^{14}$C lie at energies above 6 MeV, and there is no 
	significant $\gamma$ line around 1.6 MeV, (ii) the $^{14}$C
	isotope was produced through projectile fragmentation reaction of
	the \Cb\ (or \Ca) in the secondary target, and as such its
	``background spectrum'' resembles that of for \Cb\ (or \Ca).

	As mentioned earlier, a key point of the present work is the
	elimination of the dependence of the measured lifetime on
	the $\gamma$-ray angular distribution.  
	For this purpose, we determined the deficiency of the
	$\gamma$-ray yields due to the lead shield.  
	The deficiency ($D^i$) of the $i$-th layer detectors is
	defined as the ratio between the yields detected 
	with and without the shield, i.e.  
\begin{equation}
\label{defdeficiency}
	D^i = f_bN^i_{\rm wpb}/N^i_{\rm wopb},
\end{equation}
        where $f_b$ is the normalization factor for different total number
	of beam particles in the two measurements.
	The lifetime was determined by comparison of the measured deficiency
	($D^i_{\rm exp}$) with the simulated one ($D^i_{\rm sim}$).

\section{Monte Carlo Simulations}
\label{sec_simulation}
	The simulated deficiency of each layer as a function of various 
	lifetime for the respective de-excitation $\gamma$ rays 
	was obtained by performing Monte Carlo simulations using
	the GEANT code~\cite{geant}.
	We have taken into account the geometry of the experimental setup,
	including the shape of the detectors, in the simulations.
	The geometry was checked by performing separate measurements in
	which $^{137}$Cs, $^{22}$Na and $^{60}$Co
	standard sources emitting 662-keV, 1275-keV, 1173-keV, and 1333-keV
	$\gamma$~rays were placed at several positions from $z=-0.15$ to
	5.15~cm.  The deficiencies of all layers measured for the respective
	target position were reproduced by the simulation within accuracies
	of $\pm$~3\% and $\pm$~7\% for layers with $D^i_{\rm exp} \ge 0.2$
	and $D^i_{\rm exp} < 0.2$, respectively.

	The simulation was then applied to the case of $\gamma$-rays
	emitted from the de-exciting particles in flight. For this
	simulation, we have considered the experimentally obtained parameters
	such as the energy and emittance of the projectile, the angular
	spread due to inelastic scattering and multiple scattering, and the
	energy loss in the target.  Apart from 
	determining the $D^i_{\rm sim}$, the response functions obtained
	from the simulation were also used for the fitting of the
	experimental $\gamma$-ray energy spectra.
	Note that the strong dependence of $D^i_{\rm sim}$ on the lifetime
	is exhibited in 
	Fig.~\ref{deficiency_18C}, Fig.~\ref{deficiency_11be},
	Fig.~\ref{deficiency_16n} and Fig.~\ref{deficiency_16C}.

\section{Results}
\label{sec_result}       

	We deduced the $D^i_{\rm exp}$ using Eq.~\ref{defdeficiency}
	and determined the lifetimes for the respective excited states.
	In the case of the inelastic scattering with the
	40-MeV/nucleon \Ca\ beam, we revised the lifetime of the
	$2^+_1$ state reported previously~\cite{imai} by incorporating the
	measured $\gamma$-ray angular distribution.  The results from
	these measurements are presented.

\subsection{Lifetime of $2^+_1$ state in $^{18}$C}

\begin{figure}
\begin{center}
\includegraphics[width=80mm]{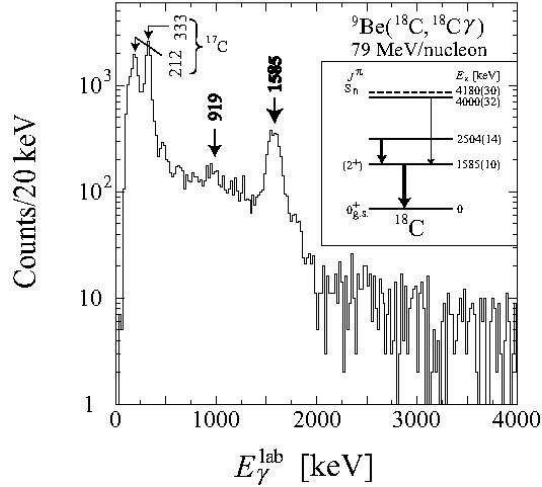}
\end{center}
\caption{ $\gamma$-ray energy spectrum for the inelastic scattering of
  the 79-MeV/nucleon \Cb\ nuclei on $^9$Be.
  The $\gamma$ rays were detected by the L6 NaI(Tl) detectors.  The inset
  shows the energy level scheme of $^{18}$C and the known $\gamma$-ray
  transitions~\cite{18clevel}.}
\label{espectrum_18C}
\end{figure}

        We show in Fig.~\ref{espectrum_18C} the
	$\gamma$-ray energy spectrum measured with the L6 NaI(Tl) detectors,
	which were located at polar angle of about 90$^{\circ}$ in the
	center of mass frame.  The spectrum was obtained with the center
	setup and without the lead shield.  The significant peak around
	1500~keV corresponds to the 1585-keV line from the
	$2^+_1 \rightarrow 0^+_{\rm gs}$ transition, while the minor peak
	around 1000~keV corresponds to the 919-keV line from the transition
	between the 2504-keV excited state and the $2^+_1$ state.
        The level scheme for \Cb\ as shown in the inset of
	Fig.~\ref{espectrum_18C} has been proposed recently based on the
	in-beam $\gamma$-spectroscopy~\cite{18clevel}.
	The two $\gamma$ lines observed correspond to the
	transitions shown by the bold arrows in the level scheme.
	The two significant peaks around 200 keV and 300 keV are from
	the known transitions in $^{17}$C~\cite{18clevel,elekes2,dsuzuki};
	the $\gamma$-ray energies shown are taken from Ref.~\cite{dsuzuki}.

	By considering the sum spectrum of all detectors, the cascade 
	contribution from the second excited state to the $2^+_1$ state
	was determined to be about 9\% relative to the
	$2^+_1 \rightarrow 0^+_{\rm gs}$ transition.  Since the error
	that might arise from the cascade contribution is negligibly small
	compared to the systematic error mentioned below, we have neglected
	the cascade contribution in our analysis.

\begin{figure}
\begin{center}
  \includegraphics[scale=0.55,angle=90]{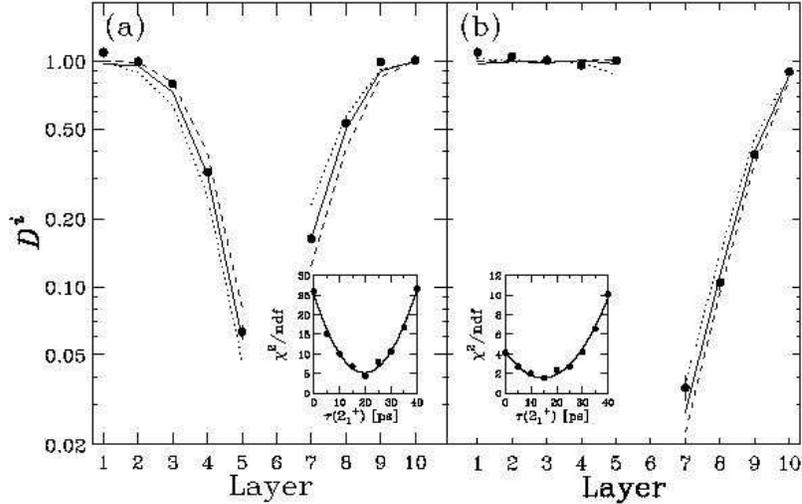}
\end{center}
\caption{ 
  The $D^i_{\rm exp}$'s, denoted by the filled circles, of the
  respective layers for the $2^+_1\rightarrow 0^+_{\rm gs}$ transition
  in $^{18}$C as compared with the simulated values.  The dashed, solid
  and dotted lines represent the $D^i_{\rm sim}$ values calculated for
  $\tau(2^+_1) = $ 0, 20, and 40~ps, respectively.  (a) The $D^i$ distribution
  for the center setup.
  (b) The $D^{i}$ distribution for the upstream setup.  The insets of the
  figures show the reduced $\chi^2$ distributions as functions of
  $\tau(2^+_1)$.}
\label{deficiency_18C}
\end{figure}

	The simulated deficiencies were obtained for several assumed
	mean lifetimes ranging from $\tau(2^+_1) = 0$~ps to 40~ps.  
	As an example, we compare the $D^i_{\rm exp}$ value with 
	the $D^i_{\rm sim}$ values obtained for $\tau(2^+_1)=$~0, 20, and
	40~ps.  The $D^i_{\rm exp}$'s (filled circles) and the
	$D^i_{\rm sim}$'s for $\tau(2^+_1)=$~0 ps (dashed line), 20 ps
	(solid line) and 40 ps (dotted line) are plotted in
	Fig.~\ref{deficiency_18C} for the (a) center and (b) upstream
	setups.  Note that we have omitted the experiment data for
	the layers with $N^i_{\rm wPb} \approx 0$, which do not
	contribute to the determination of the mean lifetime.
	The figures indicate that the mean lifetime of the $2^+_1$ state
	locates around 20~ps.  The mean lifetime was determined by
	searching the $\chi^2$ minimum; the $\chi^2(\tau)$ is given by
\begin{equation}
     \chi^2(\tau) =  \sum_{i} \frac{(D^i_{\rm exp}-D^i_{\rm sim}(\tau))^2}{(\delta D^i_{\rm exp})^2},
\end{equation}
        where the $\delta D^i_{\rm exp}$ represents the statistical
        error.  The $\chi^2$ distributions for $\tau(2^+_1)$ from 0~ps
        to 40~ps, as shown in the insets of Fig.~\ref{deficiency_18C},
        give $\tau(2^+_1) = 19.9 \pm 1.0$~ps and $14.9 \pm
        2.1$ ps for the center and upstream setups, respectively.  We
        adopted the weighted mean of these two values, $18.9 \pm
        0.9$~ps, as the mean lifetime for the $2^+_1$ state in \Cb.

	The systematic error mainly composed of two uncertainties.
	One is attributed to the discrepancy between the measured
	deficiency and the simulated one, and the other is due to
	the uncertainty of the target position, which was about 0.5 mm.
	The former was estimated to be 0.5~ps by changing the
	$D^i_{\rm sim}(\tau)$ randomly within $\pm 3\%$ (or $\pm 7\%$)
	and observing the deviation of the lifetime thus obtained.
	For the latter, the uncertainty in the target position,
	which corresponds to the uncertainty in the emission point,
	resulted in an uncertainty
	of about 4.4 ps for the \Cb\ ejectiles traveling at about
	38$\%$ of the speed of light.  The resultant systematic error,
	defined as the root sum square of these two values, was determined
	to be 4.4~ps.  The mean lifetime thus obtained is $18.9 \pm
	0.9 ({\rm stat}) \pm 4.4 ({\rm syst})$~ps.  For simplicity, the
	notations for the statistical (stat) and the systematic (syst)
	errors will be omitted from now onwards.

\subsection{Lifetime of $1/2^{-}$ state in $^{11}$Be}        
\begin{figure}
\begin{center}
\includegraphics[width=80mm]{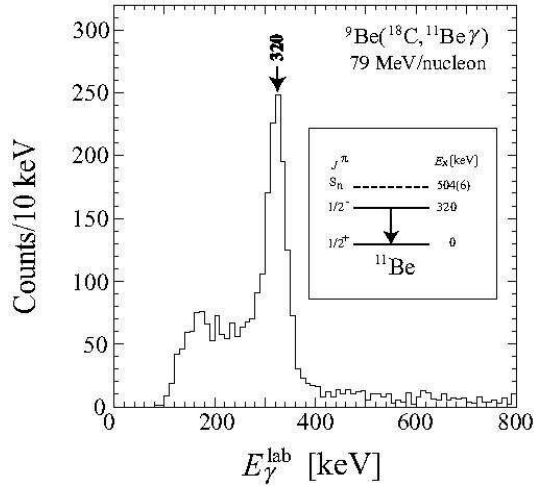}
\end{center}
\caption{ $\gamma$-ray energy spectrum in coincidence with the $^{11}$Be
  ejectiles.  The $\gamma$ rays were measured with the L6 NaI(Tl) detectors. 
  The inset in the figure presents the energy level scheme and the lifetime 
  of the first excited state of $^{11}$Be.}
\label{espectrum_11be}
\end{figure}

	Figure~\ref{espectrum_11be} shows an example of the
	$\gamma$-ray energy spectrum measured with the L6 NaI(Tl) detectors
	in coincidence with the $^{11}$Be ejectiles.
	The spectrum was obtained with the center setup and without
	the lead shield.  Only one peak that corresponds to the transition
	of the $1/2^- \rightarrow 1/2^+$ is observed at around 300~keV.
	The $1/2^-$ state is the only known bound excited
        state that decays through $\gamma$ transition as shown by the
        level scheme~\cite{ToI} in the inset of
        Fig.~\ref{espectrum_11be}.  The mean lifetime is
        $\tau(1/2^-)=$0.166(14)~ps~\cite{ToI}, which is much shorter than the
        mean lifetime of \Cb ($2^+_1$).
	By applying the RSM to this short lifetime, we verified 
	the lower limit of the dynamic range of the method.

\begin{figure}
\begin{center}
  \includegraphics[scale=0.55,angle=90]{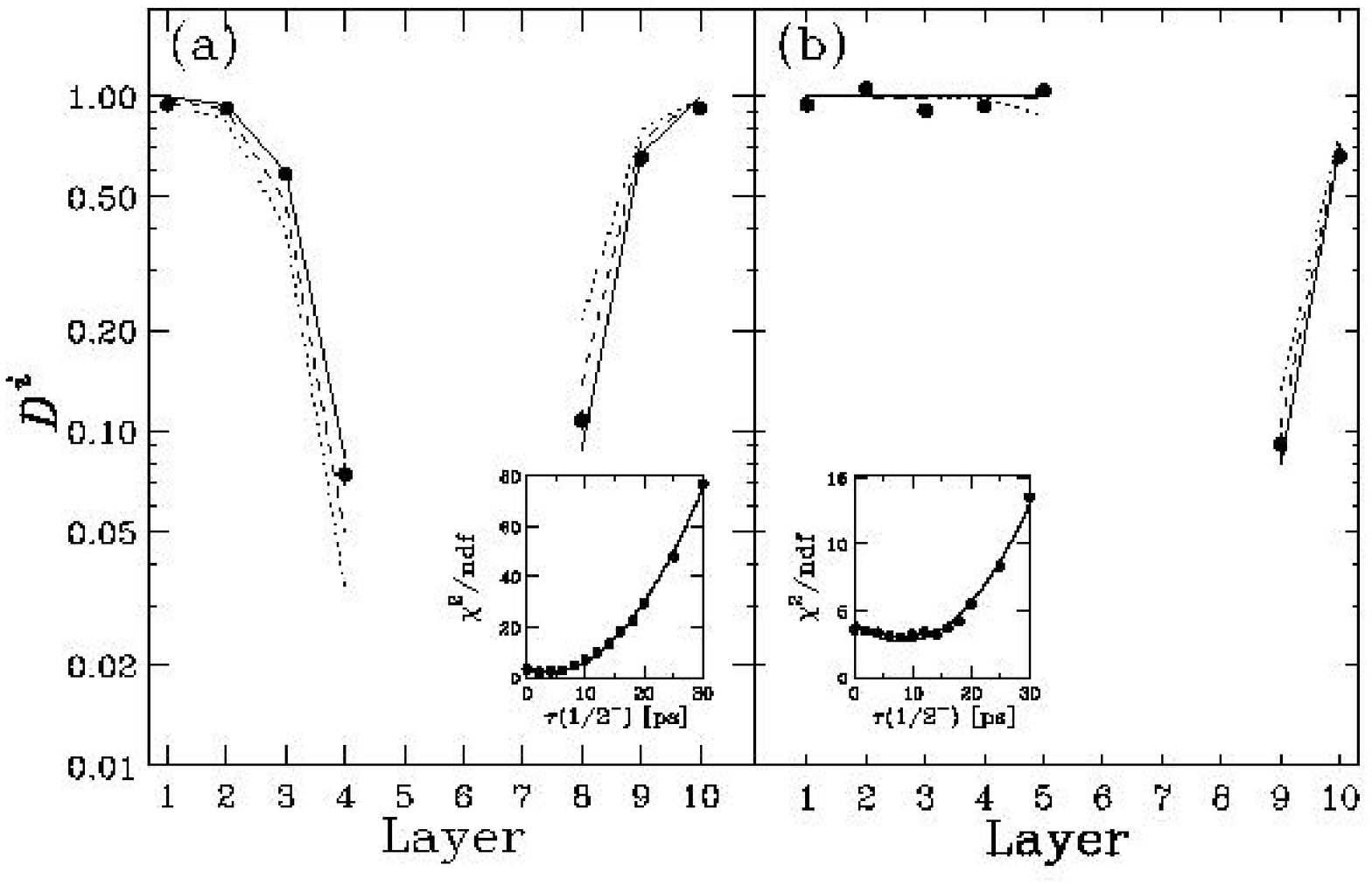}
\end{center}
\caption{ 
  The $D^i_{\rm exp}$'s, denoted by the filled circles, of the
  respective layers for the $1/2^-\rightarrow 1/2^+$ transition in $^{11}$Be
  as compared with the simulated values.  
  The solid, dashed and dotted lines represent the $D^i_{\rm sim}$
  calculated for $\tau(1/2^-) = $ 0, 15, and 30~ps,
  respectively.  (a) The $D^i$ distribution for the center setup.
  (b) The $D^{i}$ distribution for the upstream setup.  The insets show
  the reduced $\chi^2$ distributions as functions of
  $\tau(1/2^-)$.  }
\label{deficiency_11be}
\end{figure}

        The $D^i_{\rm exp}$ and the $D^i_{\rm sim}$ values for
	$\tau(1/2^-) =$~0, 15, ad 30~ps are shown in
	Fig.~\ref{deficiency_11be} for the two different target positions.
	As is obvious from the plots, the deviation of the $D^i_{\rm sim}$
	from the $D^i_{\rm exp}$ becomes larger with longer
	assumed mean lifetime.  The simulated deficiencies for $\tau(1/2^-)$
	ranging from  0~ps to 30~ps were used to obtain the $\chi ^2$ value
	as a function of $\tau(1/2^-)$.  From the $\chi^2$ distribution, the
	mean lifetimes were determined to be $3.5 \pm 1.3$~ps and
	$9.5 \pm 2.8$~ps for the center and the upstream setups, respectively.
	The weighted mean of the two values, i.e. $4.6 \pm 1.1$~ps was
	adopted.

	The sources of the systematic error are similar to those of the
	case of \Cb.  Nonetheless, it is sufficient to consider only
	the one due to the uncertainty in the target position of
	$\pm 0.5$~mm, which was dominant.  For the $^{11}$Be ejectiles
	with velocity about 37$\%$ of the speed of light, the associated
	uncertainty in the determined mean lifetime is about 4.5~ps.  Thus,
	the resultant mean lifetime for the $1/2^-$ state is
	$4.6 \pm 1.1 \pm 4.5$~ps.  
	This result is consistent with the reference value of
	0.166(14)~ps~\cite{ToI}.
	The large error shows that the RSM is not suitable for determination
	of the lifetime below 10~ps.

\subsection{Lifetime of $3^{-}$ state in $^{16}$N}
\begin{figure}
\begin{center}
\includegraphics[width=80mm]{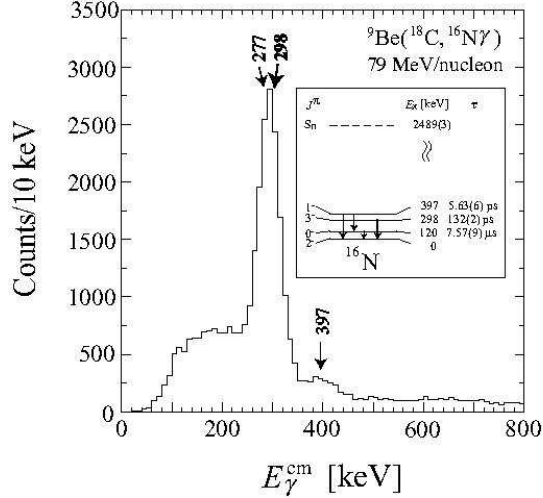}
\end{center}
\caption{ Doppler corrected $\gamma$-ray energy spectrum in coincidence
  with the $^{16}$N ejectiles.  The $\gamma$ rays were measured with all
  NaI(Tl) detectors.  The inset shows the energy level scheme
  and the known mean lifetimes of the exited states in $^{16}$N.}
\label{espectrum_16n}
\end{figure}

        The validity of the RSM was also tested by measuring the known
	lifetime of the $3^-$ state of $^{16}$N.  
	The excited $3^-$ state decays to the ground $2^-$ state with
	the mean lifetime of $\tau(3^-)=$ 132(2)~ps~\cite{ToI}.
	The Doppler-corrected $\gamma$-ray energy spectrum measured by
	all NaI(Tl) detectors is shown in Fig.~\ref{espectrum_16n}.	
	A minor and a prominent peaks are observed around 400~keV and
	300~keV, respectively.  The 400-keV peak corresponds to the
	397-keV $\gamma$ line from the $1^- \rightarrow 2^-$ transition, while
	the 300-keV peak includes two $\gamma$ lines of 277~keV and 298~keV
	from the $1^- \rightarrow 0^-$ and $3^- \rightarrow 2^-$ transitions,
	as shown by the level
	scheme~\cite{ToI} in the inset of Fig.~\ref{espectrum_16n}.

	As is evident from Fig.~\ref{espectrum_16n}, the 277-keV and
	the 298-keV peaks were not resolved in the energy spectrum.
	Hence, to determine
	the $N^i_{\rm wopb}$ and $N^i_{\rm wpb}$ for the 298-keV peak,
	it was necessary to determine the contribution of the 277-keV
	line in the 300-keV peak.  We determined the contribution by fitting
	the energy spectrum of all NaI(Tl) detectors with the simulated
	response function for the 298-keV $\gamma$ line, and a combined
	response function for the 277-keV and 397-keV lines, taking into
	account the branching ratio~\cite{ToI} of the decays from the
	$1^-$ state.  The contribution of the 277-keV line was determined
	to be about 25$\%$.  By fixing the 277-keV contribution to 25$\%$, 
	we fitted the energy spectrum for each layer
	with the simulated response functions and the $\gamma$-ray
	spectrum for the $^{14}$C isotope as background.

\begin{figure}
\begin{center}
  \includegraphics[scale=0.55,angle=90]{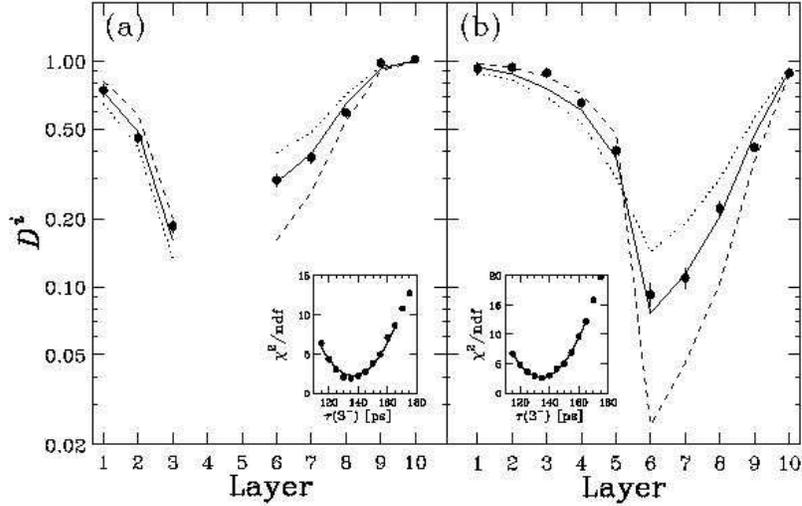}
\end{center}
\caption{ 
  The $D^i_{\rm exp}$'s, denoted by the filled circles, of the respective
  layers for the $3^-\rightarrow 2^-$ transition in $^{16}$N as compared
  with the simulated values.  The dashed,
  solid and dotted lines represent the $D^i_{\rm sim}$ calculated for
  $\tau(3^-) = $ 100, 140, and 180~ps, respectively.
  (a) The $D^i$ distribution for the center setup.
  (b) The $D^i$ distribution for the upstream setup.  The insets show
  the reduced $\chi^2$ distributions as functions of $\tau(3^-)$.  }
\label{deficiency_16n}
\end{figure}

	Figure~\ref{deficiency_16n} shows the $D^i_{\rm exp}$ and the
	$D^i_{\rm sim}$ values simulated for $\tau(3^-) =$~100, 140 and 180~ps
	for the two target positions.  The plots clearly indicate that
	the lifetime locates around 140~ps.  The simulated deficiencies
	for $\tau(3^-)$ ranging from 115~ps to 175~ps were used to obtain
	the $\chi ^2$ values as a function of $\tau(3^-)$.  From the $\chi^2$
	distributions, we obtained $\tau(3^-) = 137 \pm 4$~ps and
	$136 \pm 3$ ps for the center and the upstream setups, respectively.
	The weighted mean of the two values thus obtained is $136 \pm 3$~ps.

	Unlike the cases for \Cb\ and $^{11}$Be, the systematic error is
	mainly due to the uncertainty in the determination of the
	277-keV contribution in the 300-keV peak.  This uncertainty,
	determined to be about $5\%$, is ascribed to the uncertainty in
	the energy calibration of the NaI(Tl) detectors, which was about
	2 keV at 300 keV.  The resultant systematic error
	distributes from $-7$~ps to $+ 11$~ps.  Hence, the mean lifetime
	for the $3^-$ state was determined to be
	$136 \pm 3 ^{+11}_{-7}$~ps, which is in
	good agreement with the reference value of 132(2)~ps~\cite{ToI}.

	We shall note that the energies of the $\gamma$ rays emitted by
	$^{11}$Be and $^{16}$N are well below that of from the $2^+_1$ state
	in \Cb, which is in the region of 1300 -- 2300 keV depending on
	the layer.  To rule out any possible energy dependence, we have
	also measured the mean lifetime of the $2^+_1$ state in $^{12}$B.
	The result is consistent with the value of 260 fs~\cite{ToI}
	given in the literature.

\subsection{Lifetime of $2^+_1$ state in $^{16}$C revisited}

\subsubsection{Inelastic scattering of 72-MeV/nucleon \Ca\ beam}

\begin{figure}
\begin{center}
\includegraphics[width=87mm]{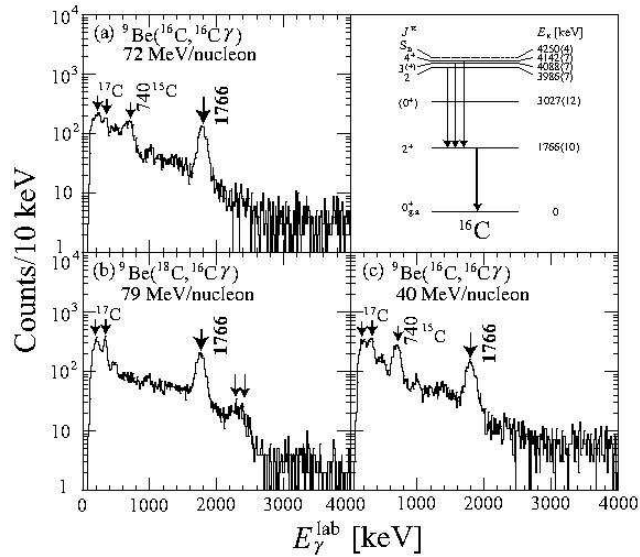}
\end{center}
\caption{ $\gamma$-ray energy spectra for \Ca\ obtained with the L6
  NaI(Tl) detectors during the measurement with the center setup and
  without the lead shield.
  (a) The inelastic scattering of $^{16}$C at 72~MeV/nucleon.
  (b) The breakup reaction of $^{18}$C to $^{16}$C at 79~MeV/nucleon.
  (c) The inelastic scattering of $^{16}$C at 40~MeV/nucleon.
  The level scheme with the known transitions in \Ca\ is shown in
  the top right panel.
}
\label{espectra_16C}
\end{figure}

	Figure~\ref{espectra_16C}~(a) shows an example
	of the $\gamma$-ray energy spectrum in coincidence with the \Ca\
	ejectiles obtained with the L6 NaI(Tl)
	detectors during the measurement with the center setup and without
	the lead shield.  The significant peak around 1800~keV
	corresponds to the 1766-keV $\gamma$ line from the
	$2^+_1 \rightarrow 0^+_{\rm gs}$ transition.  The two peaks
	around 300 keV and the peak around 740 keV are from the
	known transitions in $^{17}$C~\cite{18clevel,dsuzuki,elekes2} and
	$^{15}$C~\cite{ToI}, respectively.  The minor peak around
	1000 keV, which was also observed at the same energy in the
	spectra of the other layers as well as in the spectra of other
	reaction channels, is likely to correspond to
	the 980.7-keV~\cite{ToI} line from $^8$Li produced through
	fragmentation of the $^9$Be reaction target.  
	Although no notable peak can be observed around 2300~keV in
	the figure, the sum spectrum of all NaI(Tl) detectors
	exhibits a small peak,  which corresponds to the cascade
	transition from the higher excited state(s) as shown by
	the energy level scheme~\cite{ToI} in the top right panel of
	Fig.~\ref{espectra_16C}.  These higher excited states cannot be
	identified due to the resolution of the NaI(Tl) detectors.
	Nonetheless, by fitting the sum spectrum with a simulated
	response function for 2300-keV $\gamma$ rays, the cascade
	contribution was determined to be 9\% relative to the
	$2^+_1 \rightarrow 0^+_{\rm gs}$ transition.
	Similar to the case for \Cb, this cascade contribution is negligibly
	small, and thus was not considered in our analysis.

\begin{figure}
\begin{center}
  \includegraphics[scale=0.55,angle=90]{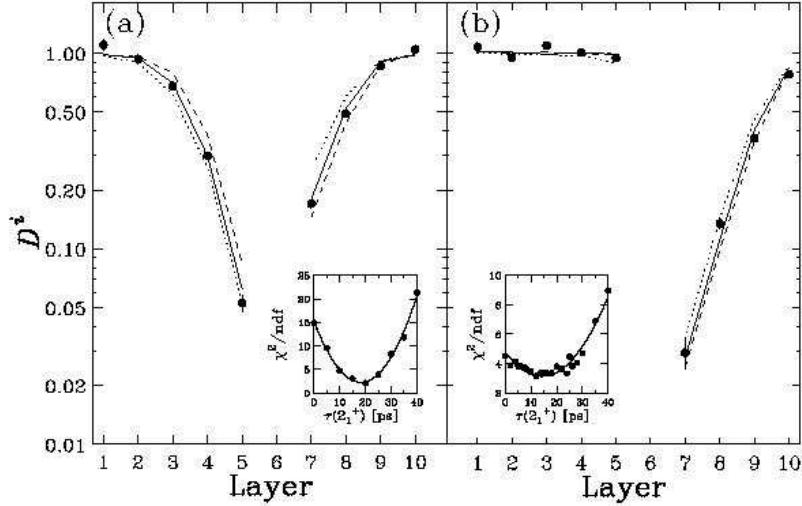}
\end{center}
\caption{ 
  Results for the inelastic scattering of $72$-MeV/nucleon \Ca\ beam.
  The $D^i_{\rm exp}$'s, denoted by the filled circles, 
  of the respective layers for the $2^+_1\rightarrow 0^+_{\rm gs}$
  transition in $^{16}$C as compared with the simulated values.
  The dashed, solid and dotted lines represent the $D^i_{\rm sim}$
  calculated for $\tau(2^+_1) = $ 0, 20, and 40~ps, respectively.
  (a) The $D^i$ distribution for the center setup.
  (b) The $D^i$ distribution for the upstream setup.
  The insets show the reduced $\chi^2$ distributions as functions of
  $\tau(2^+_1)$.  }
 \label{deficiency_16C}
\end{figure}

	The experimental deficiency of each layer is compared with the
	simulated ones of $\tau(2^+_1) =$~0, 20, and 40~ps in
	Fig.~\ref{deficiency_16C} for
	the two target positions.  The simulated curves ranging from
	$\tau(2^+_1) = 0$~ps to 40~ps were used to obtain the $\chi
	^2$ values as a function of $\tau(2^+_1)$.  The $\chi^2$
	distributions give $\tau(2^+_1) = 18.3 \pm
	1.8$~ps and $14.8 \pm 3.9$ ps for the center and the upstream
	setups, respectively.  Taking the weighted mean of the two
	values, the mean lifetime was determined to be $17.7 \pm 1.6$~ps.
	This value is about four times shorter than the value reported
	previously~\cite{imai}.

	The sources of the systematic error are similar to those 
	for \Cb.  Based on the same analysis, the systematic errors
	attributed to the uncertainties of the $D^i_{\rm sim}$ and the
	uncertainty of the target position were determined to be 0.5~ps and
	4.6~ps, respectively.  Adopting the root sum square of these
	two values as the systematic error, the mean lifetime
	thus obtained is 
	$17.6 \pm 1.6 \pm 4.6$~ps.

\subsubsection{Breakup reaction of 79-MeV/nucleon \Cb\ beam}

	Figure~\ref{espectra_16C}~(b) shows the $\gamma$ ray
	energy spectrum measured by the L6 NaI(Tl) detectors.  The spectrum
	was obtained with the center setup and without the
	lead shield.  A significant peak around 1800~keV corresponds
	to the $2^+_1 \rightarrow 0^+_{\rm gs}$ transition of 1766~keV.
	Compared with the energy spectrum of the inelastic scattering, the
	cascade transition observed around 2300~keV was enhanced.  By
	considering the sum spectrum for all NaI(Tl) detectors, the
	cascade contribution was determined to be 22\% relative to the
	$2^+_1 \rightarrow 0^+_{\rm gs}$ transition.

	In contrast to the case for inelastic channel, the cascade
	contribution was quite large.  A possible sizable lifetime for
	the higher excited state(s) might affect the outcome of the lifetime
	determination for the $2^+_1$ state. Hence, to determine the
	$\tau(2^+_1)$ and simultaneously take into account the cascade
	contribution, we carried out a minimization two-parameter $\chi^2$
	analysis.  The $\chi^2$ values were obtained with the measured
	$D^i_{\rm exp}$'s and the $D^i_{\rm sim}$'s.  Here,
	the $D^i_{\rm sim}$'s were obtained for $\tau(2^+_1)$ and the lifetime
	of the higher excited state, denoted by $\tau({\rm cascade})$
	hereinafter, from 0~ps to 40~ps.  A minimum was observed
	around $\tau(2^+_1) = 20$ ps and $\tau({\rm cascade}) = 0$ ps.
	The mean lifetimes for the $2^+_1$ state were determined to be
	$20.2 \pm 8.7$~ps and $16.9 \pm 16.8$ ps for the center and
	the upstream setups, respectively.  Taking the weighted mean of
	the two values, we obtained $\tau(2^+_1) = 19.5 \pm 7.7$~ps.

	The sources of the systematic error are similar to the ones for
	the inelastic channel.  The error attributed to the uncertainties of
	the $D^i_{\rm sim}$'s is 0.5~ps.  Meanwhile, the speed of the \Ca\
	nuclei was about 37$\%$ of the speed of light.  Hence, the systematic
	error ascribed to the uncertainty in the target position was about
	4.5~ps.  Taking these two systematic errors into consideration,
	the resultant mean lifetime for the breakup channel becomes 
	$19.5 \pm 7.7 \pm 4.5$~ps.  This value is consistent with the
	value determined from the inelastic channel.

\subsubsection{Inelastic scattering of 40-MeV/nucleon \Ca\ beam}

	The de-excitation $\gamma$ rays from the \Ca\ nuclei 
	inelastically excited with the Be target were measured at
	40~MeV/nucleon.  Figure~\ref{espectra_16C}~(c) shows the
	$\gamma$-ray energy spectrum measured with the L6 NaI(Tl) detectors.
	The spectrum was obtained with the upstream setup and
	without the lead shield.  Since no notable peak was observed around
	2300~keV, we have neglected the effect of the cascade contribution
	on the angular distribution of the $2^+_1 \rightarrow 0^+_{\rm gs}$
	transition.

        In the previous experiment~\cite{imai}, we determined the
	$\tau(2^+_1)$ by comparing the relative yields of de-excitation
	$\gamma$ rays measured with two layers of NaI(Tl) detectors 
	with a simulation assuming an angular distribution of $\gamma$ rays.
	The layers, labeled R1 and R2, were located just upstream
	of the target with central angles of 135$^{\circ}$
	and 116$^{\circ}$ in the center of mass frame, respectively.

\begin{figure}
\begin{center}
\includegraphics[width=80mm]{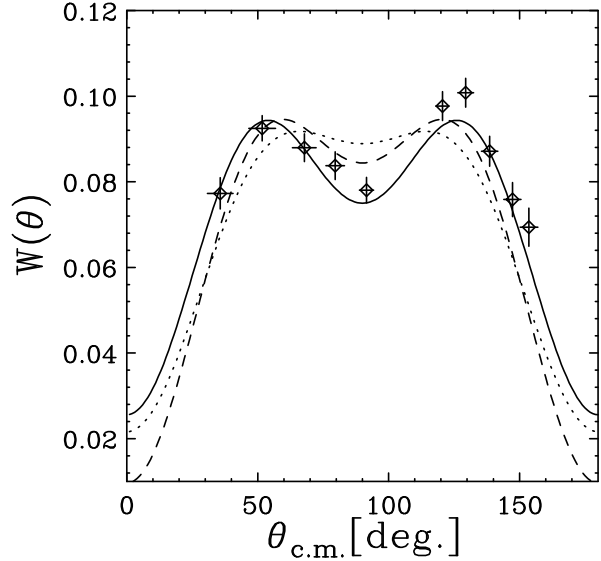}
\end{center}
\caption{ The angular distribution of the de-excitation $\gamma$ rays
  from the $2^+_1$ state in \Ca.  The solid line indicates the best
  fitted angular distribution.  The dashed and dotted lines represent the
  calculated distributions for OM1 and OM2, respectively.  }
\label{angdist_16C}
\end{figure}

        The measured angular distribution of $\gamma$ rays emitted
	from the $2^+_1$ state is shown in Fig.~\ref{angdist_16C}, together
	with the calculated distributions obtained with the
	ECIS79~\cite{ecis79} code using two optical-model parameter sets,
	OM1~\cite{OM1} and OM2~\cite{OM2}.  These calculated distributions
	were used to determine the $\tau(2^+_1)$ in the previous work.
	From the figure, it is obvious that the calculations fail to
	reproduce the experimental data, especially at around 90$^{\circ}$.
	In order to determine the angular distribution quantitatively,
	we fitted the data with the following function:
\begin{equation}
 W(\theta) = 1/(4\pi)(1 + a  P_{2}(\cos\theta) + b
	P_{4}(\cos\theta)).
\end{equation}
	The $P_{l;(l=2,4)}$ in the above function are the Legendre
	polynomials, while the $a$ and $b$ are the coefficients.
	The best-fitted result is shown in Fig.~\ref{angdist_16C} by
	the solid curve.

\begin{figure}
\begin{center}
\includegraphics[width=80mm]{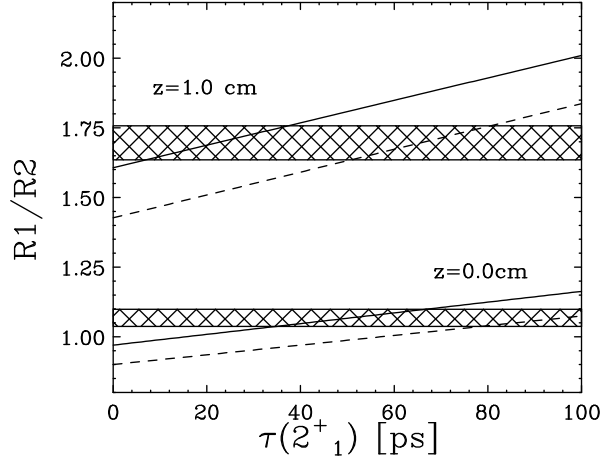}
\end{center}
\caption{ Two solid lines represent $\tau(2^+_1)$ vs. R1/R2 curves obtained
	by Monte Carlo simulation including the experimental angular
	distribution of de-excitation $\gamma$ ray for target
	positions of $z=0.0$ and 1.0~cm.  The dashed lines represent
	the original simulated R1/R2 ratio~\cite{imai}.  The hatched zones
	represent the experimentally determined R1/R2 ratios~\cite{imai}
	for the two target positions.  }
\label{newlifetime_16C}
\end{figure}

	The measured $\gamma$-ray distribution gives
	$W(135^{\circ})$/ $ W(116^{\circ})$ = $1.00\pm 0.01$.  On the
	other hand, the calculated distributions yield
	$W(135^{\circ})/W(116^{\circ})=0.91$ for OM1 and 0.88 for OM2,
	which made the simulated R1/R2 ratio smaller.  As a result, a
	longer mean lifetime was deduced.
	Figure~\ref{newlifetime_16C} shows the simulated R1/R2 ratio
	as a function of the $\tau(2^+_1)$ by incorporating the
	experimental angular distribution.  The dashed lines represent
	the original simulated R1/R2 ratio. The R1/R2 ratio measured
	in the previous work for the two target positions, $z=0.0$ and
	$1.0$~cm, are shown
	by the hatched zones.  The overlapped region between the
	experimental R1/R2 ratio and the simulated lines corresponds
	to $51 \pm 21$~ps and $20 \pm 18 $~ps for $z=0.0, 1.0$~cm,
	respectively.  The resultant mean lifetime of $34 \pm 14 $ was
	obtained by taking the weighted mean of these two values.  The
	systematic uncertainty, attributed mainly to the geometrical
	uncertainty, was estimated to be 20\%.  Hence, the revised
	lifetime is $34 \pm 14 \pm 7$~ps.
	This value is shorter than the previously reported value of
	$77 \pm 14 \pm 19$~ps~\cite{imai}, but is consistent
	with the other two experimental results given above.

\section{Discussion}
\label{sec_discuss}

\begin{table}	
\caption{Summary of the mean lifetimes of the $2^+_1$ states in \Ca\ and \Cb,
  and the corresponding $B(E2)$ values.  The superscripts $a$,$b$ and $c$
  indicate inelastic channel at 72 MeV/nucleon, breakup channel at
  79 MeV/nucleon, and inelastic channel at 40 MeV/nucleon.}
\label{table_lifetime}
\begin{tabular}{p{2cm}p{3.0cm}{l}p{3.0cm}{l}} \hline \hline
         &    $\tau$($2^+_1$) [ps] &  $B(E2)$ [$e^2$fm$^4$]      \\ \hline
\Cb      &    $18.9\pm 0.9 \pm 4.4$        &  $4.3\pm 0.2 \pm 1.0$ \\          
\Ca$^a$  &    $17.7\pm 1.6 \pm 4.6$       &  $2.7\pm 0.2 \pm 0.7$  \\
\Ca$^b$  &    $19.5\pm 7.7 \pm 4.5$       &  $2.4\pm 0.9 \pm 0.6$  \\
\Ca$^c$  &    $34 \pm 14 \pm 7$          &  $1.4\pm 0.6 \pm 0.3$  \\ \hline \hline
\end{tabular}
\end{table}

\begin{figure}
\begin{center}
\includegraphics[width=80mm]{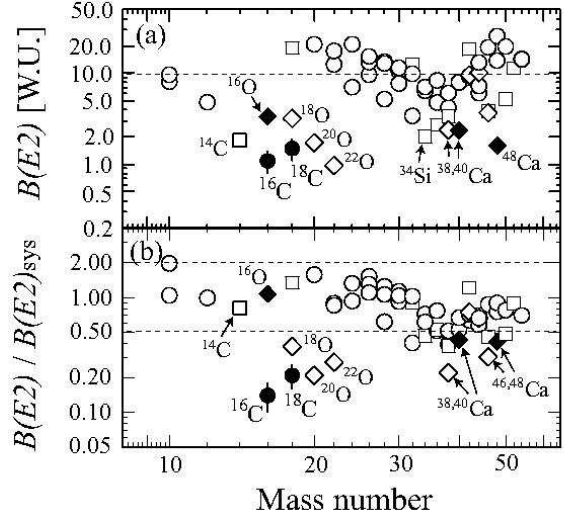}
\end{center}
\caption{(a) \BE\ values in W.u. and (b) ratios between
	the experimental \BE\ values and $B(E2)_{\rm sys}$ calculated
	using the empirical formula~\cite{raman} for even-even nuclei
	with $A \le 50$.  The filled circles denote the values of $^{16,18}$C,
	and the open circles represent data for other open-shell nuclei.
	The open squares and open diamonds denote the proton- and
	neutron-closed-shell nuclei, while the filled diamonds represent
	the double magic nuclei.  The dashed lines are intended as
	eye guides.}
\label{be2all}
\end{figure}

        The obtained $\tau(2^+_1)$ for $^{16,18}$C are summarized in
        Table~\ref{table_lifetime}.  In the case of $^{16}$C, all three
	$\tau(2^+_1)$ values obtained from the three measurements are in
	consistent with each other.  We adopt the weighted mean of the
	three values by considering only the statistical errors.
	For the systematic errors, since they are comparable in all three
	cases, we adopt the largest value (26$\%$) as the resultant
	systematic error.  The $\tau(2^+_1)$ value for \Ca\ thus obtained is 
	$18.0\pm 1.6 \pm 4.7$ ps.
	
        The $\tau(2^+_1)$'s determined in the present work correspond
	to \BE~values of $2.6\pm 0.2 \pm 0.7 \ e^2$fm$^4$ and 
	$4.3\pm 0.2 \pm 1.0 \ e^2$fm$^4$ for
	$^{16,18}$C, respectively.  Although the \BE\ value for \Cb\
	is almost twice as large as that of \Ca, it is comparable to the
	$B(E2)= 3.7$~$e^2$fm$^4$ of the closed-shell $^{14}$C nucleus.
	Both the energy of the $2^+_1$ state and the \BE\ value remain
	small in \Cb, clearly indicating that the phenomenon of
	hindered $E2$ strength observed in \Ca\ persists in \Cb.

	In Fig.~\ref{be2all}~(a), the \BE\ values obtained for
	$^{16,18}$C are compared with all known \BE\ values 
	for the even-even nuclei with $A<50$~\cite{raman2}. Nuclei
	with open shells tend to have \BE\ values greater than 10
	W.u., whereas nuclei with neutron- or/and proton shell closure 
	tend to have distinctly smaller \BE\ values. Typical
	examples of the latter category are the doubly magic nuclei,
	$^{16}$O and $^{48}$Ca, for which the \BE\ values are known to
	be 3.17 and 1.58~W.u., respectively.  The present \BE\ values
	for $^{16,18}$C are 1.1 and 1.5~W.u., respectively, which are
	even more suppressed than those of the doubly magic nuclei 
	although they are supposed to be open-shell nuclei.

	The strong hindrance of the $^{16,18}$C transition can also be
	illustrated through comparison with an empirical formula based
	on a liquid-drop model~\cite{raman}. 
	The empirical formula can be expressed by
	\[ B(E2)_{\rm sys} = 6.47 \times Z^{2}A^{-0.69}E(2^+_1)^{-1}. \]
	The experimental \BE\ values relative to $B(E2)_{\rm sys}$ are
	plotted in Fig.~\ref{be2all}(b). As noted in
	Ref.~\cite{raman2}, the \BE/$B(E2)_{\rm sys}$ ratios for
	most of the open-shell nuclei fall around 1.0, being confined
	between 0.5 and 2.0.  Even for the closed-shell nuclei, the
	ratio remain larger than 0.20.  Thus, the ratios of 0.14 and
	0.21 for $^{16,18}$C are exceptionally small. 
	In particular, the ratio for $^{14}$C with
	$E(2^+_1) = 7012$ keV~\cite{ToI}
	is as large as 0.68, suggesting different mechanisms for the small
	\BE\ in $^{14}$C and the suppression of the \BE\ values in
	$^{16,18}$C.  As in the case of \Ca~\cite{elekes,onghjin}, the
	observation of the small \BE\ value in \Cb\ despite
	the lowering of the $E(2^+_1)$ may imply a neutron-dominant
	quadrupole collectivity in \Cb.

\begin{figure}
\begin{center}
\includegraphics[width=80mm]{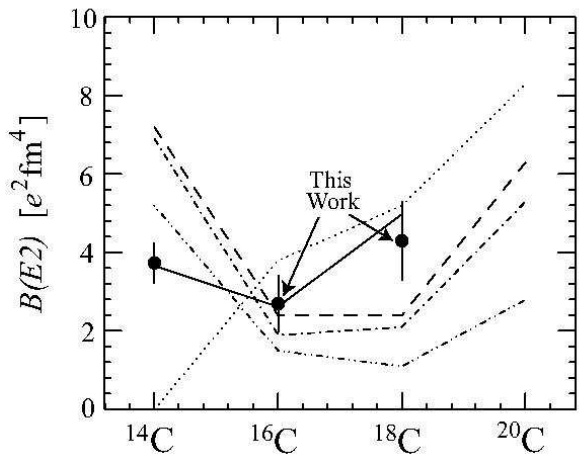}
\end{center}
\caption{Experimental \BE\ for $^{14-20}$C isotopes (filled circles)
  in comparison with the \BE\ predicted by theoretical calculations
  with the shell model~\cite{fuji}~(dashed line),
  the AMD~\cite{amd2} (dash-dotted line),
  the AMD+MSD~\cite{amd3} (dash-double-dotted line),
  the deformed Skyrme Hartree-Fock~\cite{meanfield}
  (dotted line), and 
  the ``no-core'' shell model~\cite{sfuji}~(solid line).  }
\label{be2theory}
\end{figure}

	Regarding the suppressed \BE\ values for $^{16,18}$C, 
	calculations have been performed in several theoretical frameworks.
	Figure~\ref{be2theory} shows the experimental and some of
	the theoretical \BE\ values for carbon isotopes from $^{14}$C to
	$^{20}$C.

	Calculations using the antisymmetrized molecular dynamics
	(AMD)~\cite{amd1,amd2} and the multi-Slater determinant AMD
	(AMD+MSD)~\cite{amd3} have reproduced the trend of the
	small \BE\ values in $^{16,18}$C.  In the framework of the AMD
	calculation, the presence of opposite deformations in the
	neutron and proton are said to be accounted for the
	small \BE\ in the $^{16,18}$C isotopes.  Meanwhile, another
	calculation using the deformed Hartree-Fock wave
	function~\cite{meanfield} has
	reproduced the \BE\ values in $^{16,18}$C rather well, although
	it fails to reproduce the one in the neutron-closed-shell nucleus
	$^{14}$C.

        The shell model calculations~\cite{fuji,sfuji} have also
        predicted suppressed \BE\ values for $^{16,18}$C.  In the case
        of the calculation using a Hamiltonian with the strengthened
        tensor interaction~\cite{fuji}, the energy gap between $(\pi
        0p_{1/2})$-$(\pi 0p_{3/2})$ becomes as large as the gap for
        $(\pi 1s_{1/2})$-$(\pi 0p_{1/2})$, when the neutron number
        is 8 and 10.  This large gap reduces the proton 
        matrix element.  As a result, the $0^+_{\rm gs} \rightarrow 2^+_1$
	transition is mainly contributed by the neutrons.  On the
	other hand, the dominant configuration of the valence neutrons
	in the $sd$-shell is $1s_{1/2}$, and since the valence neutrons
	in the $s$-orbit spread out widely, the effective charges
	become small.  In fact, the effective charge of neutron has been
	calculated to be $e_n =0.2e$~\cite{sagawa-asahi}, which is smaller
	than the standard value $e_n=0.5e$ used for the $sd$-shell nuclei.
	The other shell model calculation of the no-core type has reproduced
        successfully the present \BE\ values for $^{16,18}$C with a
        small neutron effective charge of $0.164e$~\cite{sfuji}.
	In the calculation, a quenched proton transition is expected.  The
        occupation numbers of proton in the $0p_{3/2}$ orbits change
        by less than 1\% for the ground and the $2^+_1$ states in
        $^{16}$C~\cite{sfuji} and $^{18}$C~\cite{sfuji2}.
	Thus, these two calculations give the same picture
        for the suppressed \BE\ value.  Note that the former shell
        model calculation predicts that the energy gap of $(\pi
        p_{1/2})$-$(\pi p_{3/2})$ decreases from $^{16}$C to $^{22}$C.
        As a result, proton matrix element will be enhanced, leading 
	to larger \BE\ values for $^{20,22}$C.

	While the above theoretical models offer possible
	interpretation for the hindered $E2$ transitions observed in
	$^{16,18}$C, more theoretical and 
	experimental studies on the ground states and other excited states,
	e.g. $2^+_1$ state in $^{20}$C, are necessary for a more 
	unified and complete understanding of the structure of the
	neutron-rich carbon isotopes.

        Finally, we comment on the experimental results of \Ca\
        reported in Ref.~\cite{elekes,onghjin}.  A smaller \BE\ value
        consistent with the one in Ref.~\cite{imai} was reported in
        Ref.~\cite{elekes}.  The underestimation of the \BE\ value in
        Ref.~\cite{elekes} may indicate the need for a microscopic
	approach in analyzing the reaction data.  In fact, a subsequent
        analysis~\cite{takashina} using the AMD wave functions~\cite{amd2}
	in the microscopic coupled-channels calculations has
	indicated a larger \BE\ value of 1.9 $e^2$fm$^4$.  
        As for the work on the inelastic proton scattering on 
        \Ca\ \cite{onghjin}, the quadrupole deformation length was
	extracted from the experiment data.  Using this deformation length
	and the \BE\ value in Ref.~\cite{imai}, the ratio of the neutron and
	proton quadrupole matrix elements $(M_n/M_p)/(N/Z)$ was determined to
	be $4.0 \pm 0.8$.  With the present \BE\ value, the $(M_n/M_p)/(N/Z)$
	value for \Ca\ becomes $1.9 \pm 0.4$, which is still very large and
	comparable to the value of $^{20}$O~\cite{o20}.  In fact,
	the $M_n$ value of about 11 fm$^2$ deduced for \Ca\ remains much
	larger than the $M_p$ ($=\sqrt{5 B(E2)/e^2}$) value, which was deduced
	to be about 4 fm$^2$ using the present \BE\ value.

\section{Summary}
\label{sec_summary}
        
        The lifetime of the $2^+_1$ state in $^{16,18}$C were
	successfully measured using the upgraded RSM with 10-layer
	NaI(Tl) array, incorporating the inelastic scattering and breakup
	reaction at around 75~MeV/nucleon.
	The $\gamma$-ray angular distribution for the inelastic scattering
	of $^{16}$C at 40~MeV/nucleon was also measured.  Incorporating
	this angular distribution into the measurement reported
	previously~\cite{imai}, the $\tau(2^+_1)$ of $^{16}$C was found to
	be in consistent with the present result.  The $\tau(2^+_1)$
	values for $^{16,18}$C thus determined were as long as around 20~ps,
	indicating that the anomalous suppression of \BE\ observed
	in \Ca\ persists in \Cb.  In the
	framework of shell model calculation, the suppressed \BE\ values
	can be attributed to the small effective charges and the widening of
	the energy gap between the $\pi(p_{1/2})$-$\pi(p_{3/2})$ orbitals.
	The present results, together with the small \BE\ values for
	$^{14}$C, suggests a possible proton-shell closure in the
	neutron-rich $^{14,16,18}$C nuclei.

\section*{Acknowledgment}
	We thank the RIKEN Ring Cyclotron staff for the stable
	$^{22}$Ne beam throughout the experiment.   We acknowledge fruitful
	discussions with S. Fujii.  HJO is grateful to the Japan Society
	for the Promotion of Science for scholarship.  This work
	was supported in part by Grant-in-Aid for Scientific Research
	No. 15204017 from the Ministry of Education, Culture, Sports,
	Science and Technology of Japan.

\end{document}